%% file: Yushkov_MASS2_CorrelationXmaxS38_ICRC2025_Id678_Submit_PoS442.tex
\documentclass[a4paper,11pt]{article}

\usepackage{pos}
\usepackage{xspace}
\usepackage{siunitx}
\usepackage{lineno}
\usepackage{enumitem}
\usepackage{wrapfig}
\usepackage[nameinlink,capitalize]{cleveref}

\newcommand{\eposlhcr}{\textsc{EPOS LHC-R}\xspace}
\newcommand{\qgsIII}{\textsc{QGSJet-III-01}\xspace}
\newcommand{\sibyll}[1]{\textsc{Sibyll\,#1}\xspace}
\newcommand{\sibylle}{\textsc{Sibyll\,2.3}e\xspace}
\newcommand{\sibyllstar}{\textsc{Sibyll${}^\bigstar$}\xspace}

\newcommand{\gcm}{\ensuremath{\mathrm{g\,cm^{-2}}}\xspace}
\newcommand{\arange}[2]{\ensuremath{\theta/\text{deg}\in[#1,#2]}\xspace}
\newcommand{\lgerange}[2]{\ensuremath{\lg(E/\mathrm{eV})\in[#1,#2]}\xspace}
\newcommand{\lge}{\ensuremath{\lg(E/\mathrm{eV})}\xspace}
\newcommand{\Xmax}{\ensuremath{X_{\mathrm{max}}}\xspace}
\newcommand{\XmaxNorm}{\ensuremath{\Xmax^{*}}\xspace}
\newcommand{\SdSizeKm}{\ensuremath{S(1000)}\xspace}
\newcommand{\SdSizeNorm}{\ensuremath{S^{*}_{38}}\xspace}
\newcommand{\rG}{\ensuremath{r_\mathrm{G}}\xspace}

\newcommand{\lnA}{\ensuremath{\ln A}\xspace}
\newcommand{\meanlnA}{\ensuremath{\langle\lnA\rangle}\xspace}
\newcommand{\sigmalnA}{\ensuremath{\sigma(\lnA)}\xspace}
\newcommand{\meanXmax}[1]{\ensuremath{\langle\Xmax^{#1}\rangle}\xspace}

\newcommand{\pao}{Pierre Auger Observatory\xspace}

\def\Offline{\mbox{$\overline{\textrm%
{Off}}$\hspace{.05em}\protect\raisebox{.4ex}%
{$\protect\underline{\textrm{line}}$}}\xspace}

\graphicspath{{pix_final/}}

\title{Constraints on the spread of nuclear masses in ultra-high-energy cosmic rays based on the Phase~I hybrid data from the Pierre Auger Observatory}
\ShortTitle{Spread of nuclear masses in UHECRs from the Pierre Auger Observatory data}

\author*[a]{Alexey Yushkov}

\affiliation[a]{Institute of Physics of the Czech Academy of Sciences, Na Slovance 1999/2, Prague, Czech Republic}

\onbehalf{for the Pierre Auger Collaboration$^b$}
\affiliation[b]{Observatorio Pierre Auger, Av.\ San Mart{\'\i}n Norte 304, 5613 Malarg\"ue, Argentina\\
Full author list: {\rm\url{https://www.auger.org/archive/authors_icrc_2025.html}}}

\emailAdd{spokespersons@auger.org}

\abstract{We present an analysis of the correlation between the depth
  of the maximum of air-shower profiles and the signal in
  water-Cherenkov stations in events registered simultaneously by the
  fluorescence and surface detectors of the Pierre Auger
  Observatory. The analysis enables us to place constraints on the
  spread of nuclear masses in ultra-high-energy cosmic rays with a
  minor impact from the experimental systematic uncertainties and
  uncertainties in air-shower simulations. Due to this unique feature,
  the correlation analysis has previously allowed us to exclude all
  pure and proton-helium compositions near the ankle in the cosmic-ray
  energy spectrum at $5\sigma$ confidence level. The same property makes
  the correlation analysis an effective tool for testing the
  consistency of predictions of the post-LHC hadronic interaction
  models, including their latest versions such as \eposlhcr,
  \qgsIII, \sibyllstar and \sibylle. In this work, the
  correlation analysis using the Phase~I hybrid data from the Pierre
  Auger Observatory is presented.  The analysis uses the newest
  generation of hadronic interaction models and covers an extended energy range
  around the ankle in the cosmic-ray spectrum.}

\ConferenceLogo{PoS_ICRC2025_logo.pdf}

\FullConference{%
39th International Cosmic Ray Conference (ICRC2025)\\
15 -- 24 July, 2025\\
Geneva, Switzerland}

\begin{document}
\maketitle

\section*{Introduction}

Determining the mass composition is a key effort in understanding the
origin of ultra-high-energy cosmic rays (UHECRs). The interpretation
of air-shower data in terms of the absolute values of primary masses
requires a comparison to predictions based on hadronic interaction
models, imposing related systematic uncertainties. However, the
correlation between the depth of the maximum of air-shower profiles,
\Xmax, and number of muons at the ground has been shown to provide
information about the spread of masses in the cosmic-ray beam with a
minor dependence on the hadronic
models~\cite{younk_corr_app2012}. Using the signal in water-Cherenkov
detectors (WCDs) at 1000 meters from the shower core as a proxy for
the muon number, the Pierre Auger Observatory~\cite{auger_nima2015}
has demonstrated that pure compositions, as well as all mixes
consisting only of two neighboring mass groups (proton-helium,
helium-CNO, CNO-iron) are excluded for energies near the ankle in the
energy spectrum~\cite{auger_mixed_plb2016}. In this work, we present a
preliminary update of this analysis using the Phase~I dataset of the
\pao and the latest versions of post-LHC hadronic interaction
models. Taking advantage of the minor dependence of the constraints on
the spread of primary masses on details of hadronic interactions, we
test the predictions of the post-LHC models by analyzing the energy
and zenith-angle dependencies of the observed correlation.

\section{The dataset and simulations}
\label{sec:data_mc}

The analysis is performed using events that were successfully
reconstructed with the Fluorescence Detector (FD) and Surface Detector
(SD) of the \pao during the period $(12/2004-12/2021)$. The FD event
selection is the same as in the \Xmax
analysis~\cite{fitoussi_pos2023}, except for the fiducial field of
view selection. This cut is designed to minimize the selection bias
(such as the loss of deep vertical events with \Xmax close to the
ground) on the \Xmax distributions and does not affect the quality of
individual events. In the correlation analysis, such bias is found to
be minimal, particularly because events with zenith angles below
\ang{35}, corresponding to an atmospheric depth of 1060\,\gcm at the
location of the Observatory, are excluded. The justification from the
point of view of the correlation analysis for selecting \ang{35} as
the minimum zenith angle is provided in~\cref{sec:zenith}.

To ensure an accurate estimation of the SD signal at 1000 meters from
the core, it is required that the WCD with the highest signal in an
event is surrounded by a hexagon with at least five active stations.
The maximum zenith angle is limited to \ang{60} to guarantee
reliable SD reconstruction~\cite{sdreco_jinst2021}.

After selection, in the energy range \lgerange{18.3}{19.5} used in
this work, the dataset consists of 9661 events. For the energy range
\lgerange{18.5}{19.0} discussed in our previous
publication~\cite{auger_mixed_plb2016}, approximately 2.8 times more
data is available in the current analysis.

The air-shower simulation library~\cite{santos_augerlib_icrc2023} for
proton, helium, oxygen and iron primary species has been produced
using CORSIKA~7.8010~\cite{corsika}. The Auger \Offline
framework~\cite{offline,santos_augerlib_icrc2023} has been used for
the detector simulation and event reconstruction. For the analysis
presented here, the newest versions of the post-LHC hadronic
interaction models, namely \eposlhcr~\cite{eposlhcr_pos2023},
\qgsIII~\cite{qgsjet3_i_prd2024,qgsjet3_ii_prd2024},
\sibyllstar~\cite{sibyllstar_icrc23}, and
\sibylle~\cite{sibyll23d_prd2020}, have been utilized. For more
information about these models and their predictions regarding \Xmax
and the muon shower content, see~\cite{pierog_fzu2025}.

\section{Constraints on the spread of primary masses}

The spread of masses in the primary beam is estimated using the
correlation between \Xmax and the signal in the WCDs located 1000
meters from the core, \SdSizeKm~\cite{auger_mixed_plb2016}.  To avoid
a decorrelation due to the spreads of energies and zenith angles, we
use \Xmax and \SdSizeKm scaled to a reference energy of 10~EeV, for
details, refer to~\cite{auger_mixed_plb2016}. Additionally, \SdSizeKm
is scaled to a zenith angle of $38^\circ$ as described
in~\cite{auger_sd1500_spectrum_prd2020}. The scaled variables are
referred to as \XmaxNorm and \SdSizeNorm. They represent the values of
\Xmax and \SdSizeKm that would have been observed if the shower had
arrived at a zenith angle of $38^\circ$ and an energy of 10~EeV. The
correlation between \XmaxNorm and \SdSizeNorm is assessed using a rank
correlation coefficient \rG proposed
in~\cite{rg_gideon_jasa1987}. Rank correlation coefficients are
invariant to the absolute values of \SdSizeKm and \Xmax, and therefore
remain unaffected by systematic uncertainties in the corresponding
predictions of hadronic models.

\begin{figure}
  \centering
  \includegraphics[width=0.49\textwidth]{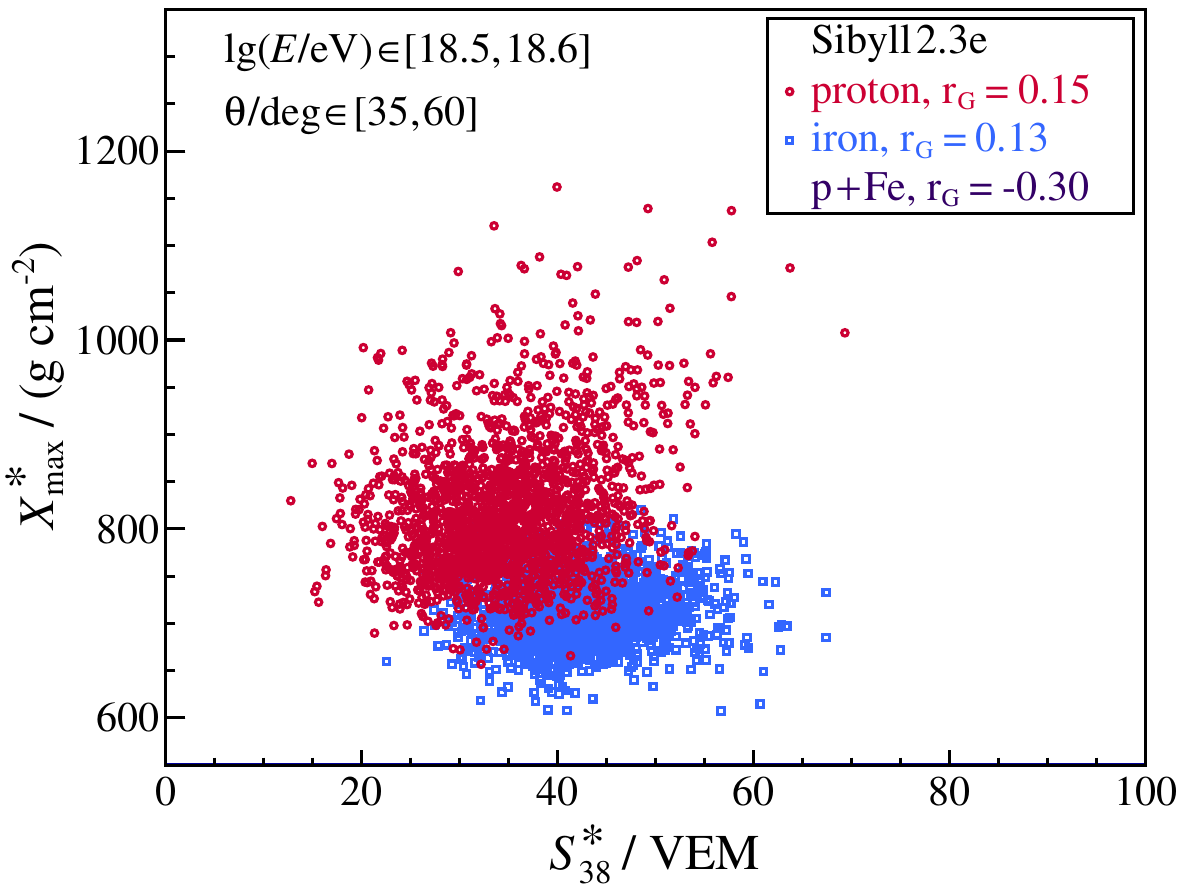}
  \includegraphics[width=0.49\textwidth]{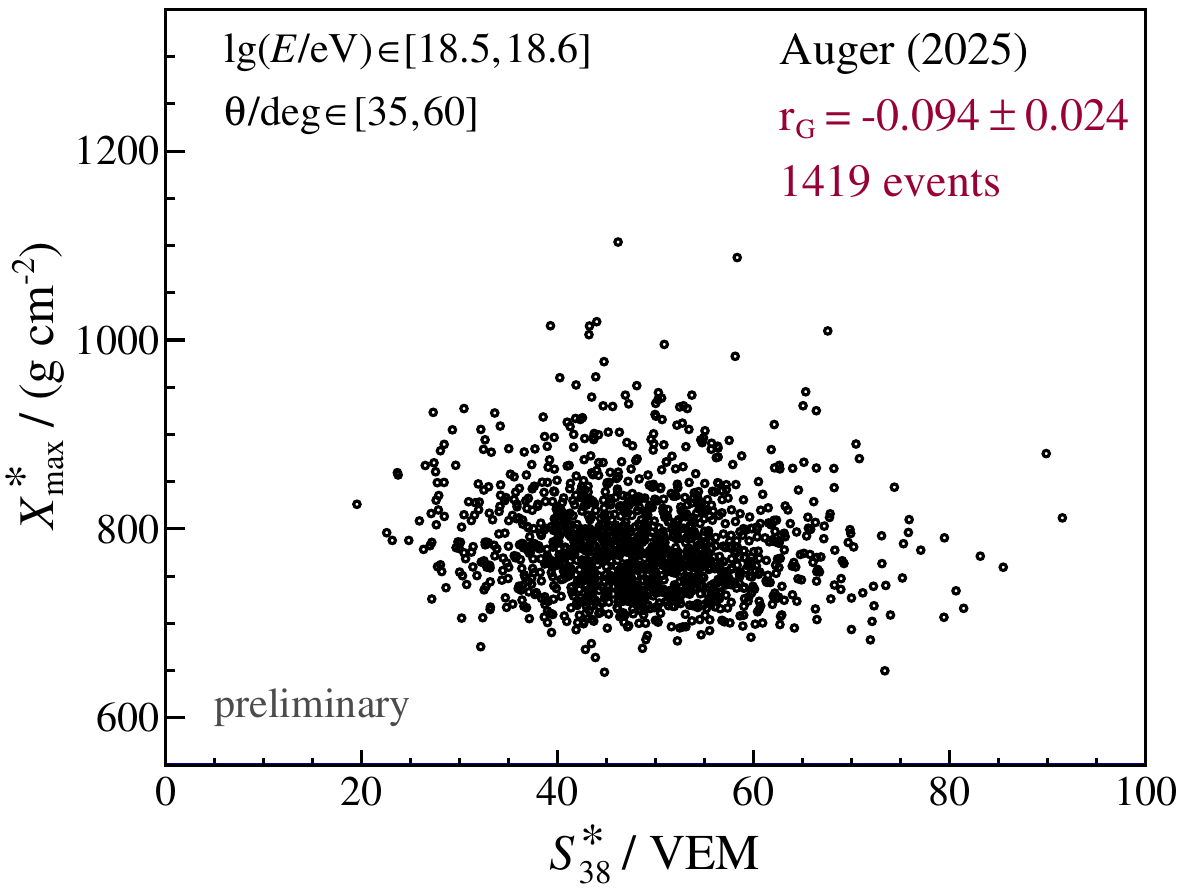}
  \caption{Correlation between \XmaxNorm and \SdSizeNorm. Left panel:
    proton and iron showers (samples of 2000 events each) simulated with \sibylle; the
    legend contains \rG values for pure beams and
    proton-iron equal mix (maximum mixing degree). Right panel: correlation in the data.
    Energy range: \lgerange{18.5}{18.6}; zenith angle range: \arange{35}{60}.}
  \label{fig:ex_scatter}
\end{figure}

The correlation analysis leverages the general characteristics of
air-shower development, which manifest as a nearly model-independent
separation between mass groups in the (\XmaxNorm, \SdSizeNorm) plane, as
well as in the nearly model-independent magnitude of the fluctuations
of these two observables. In \cref{fig:ex_scatter}, simulated
\XmaxNorm and \SdSizeNorm distributions are presented for
\sibylle proton and iron showers, alongside the corresponding
distribution observed in the data within the \lgerange{18.5}{18.6}
energy range.  The correlation is positive for pure beams but becomes
increasingly negative as the mass spread increases, reaching maximum
anticorrelation for an extreme mix of protons and iron nuclei in equal
proportions. The correlation in the data is negative, with a value in
the middle between pure beams and the extreme mix.  The statistical
uncertainty is estimated as
$\Delta\rG\simeq0.9/\sqrt{N}$~\cite{auger_mixed_plb2016}, where $N$
is the number of events in the dataset. The systematic uncertainty is
$\Delta\rG(\rm sys.)=^{+0.01}_{-0.02}$~\cite{auger_mixed_plb2016}. The
larger negative error arises from a small decorrelation introduced by
the long-term performance of the FD and SD, for which no correction
has been applied to stay conservative.

\begin{figure}
  \centering
  \includegraphics[width=0.49\textwidth]{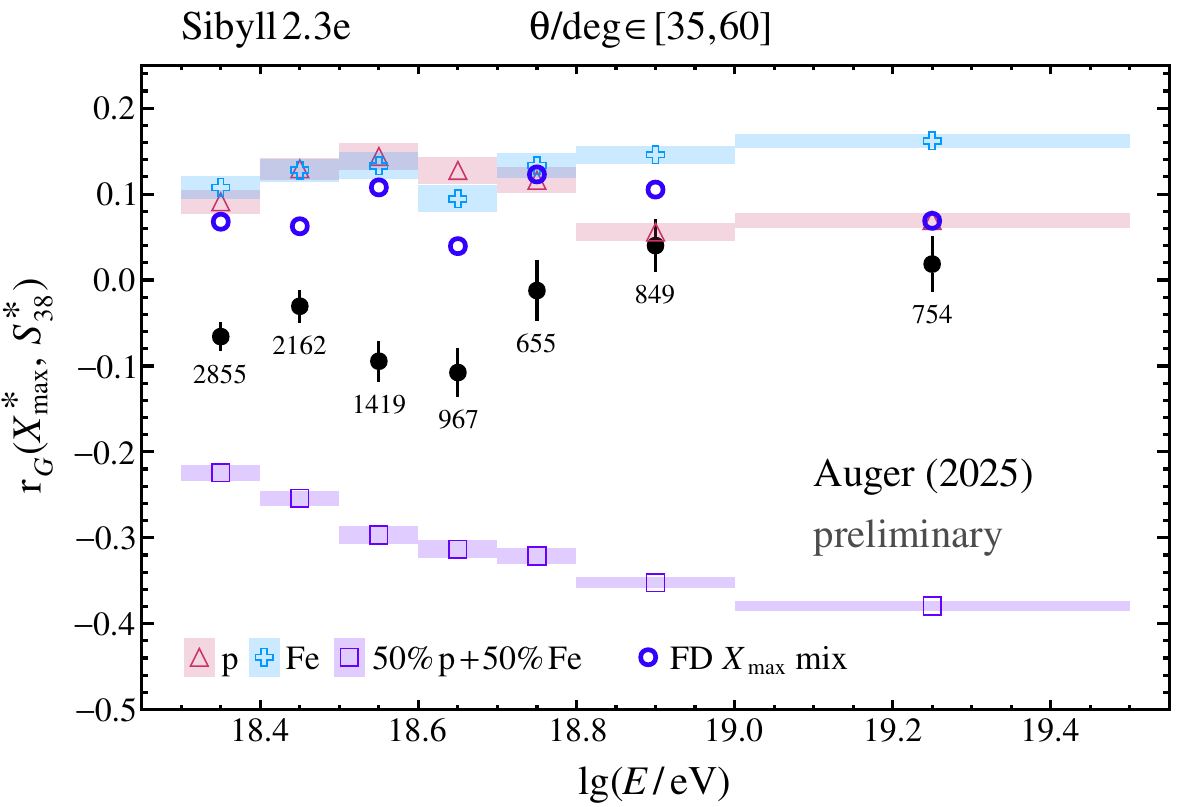}
  \includegraphics[width=0.49\textwidth]{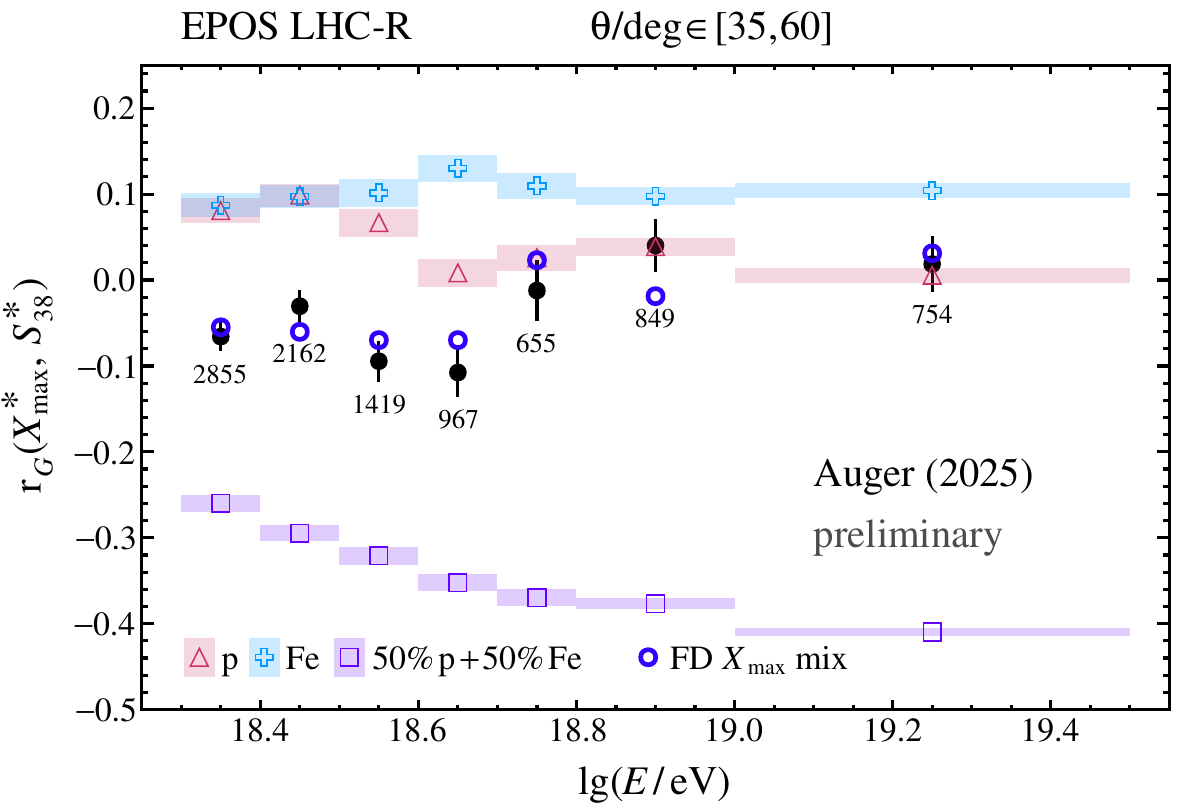}
  \caption{Energy dependence of the correlation in the data compared
    with correlations in simulations with \sibylle and \eposlhcr
    for protons, iron nuclei and proton-iron equal mix. Additionally,
    the correlation expected for the mass composition inferred from
    fits of the FD \Xmax distributions~\cite{tkachenko_icrc2023,tkachenko_icrc2025} is
    shown.}
    \label{fig:rg_energy}
\end{figure}

The energy evolution of the observed correlation is presented
in~\cref{fig:rg_energy} along with predictions from simulations using
\sibylle and \eposlhcr. Up to the ankle ($\lge\approx18.7$) the
observed correlation is negative and differs significantly from the
correlation for the pure beams. Above the ankle, \rG approaches to the
values for pure beams. These observations hold for all hadronic
interaction models used in this study. Additionally, in this plot, the
correlation expected for the mass composition obtained from the
fraction fits of the \Xmax distributions (hereafter referred to as `FD
\Xmax mix') is presented~\cite{tkachenko_icrc2023,
  tkachenko_icrc2025}. One can see that \rG for the FD \Xmax mix
differs significantly from the observed correlation below the ankle
for \sibylle (and other \textsc{Sibyll} versions\footnote{We use the
mix obtained with \sibyll{2.3d} which has the same \Xmax scale as that
of \sibylle and \sibyllstar.}), while for \eposlhcr the agreement is
good.

\begin{figure}
  \centering
  \includegraphics[width=0.49\textwidth]{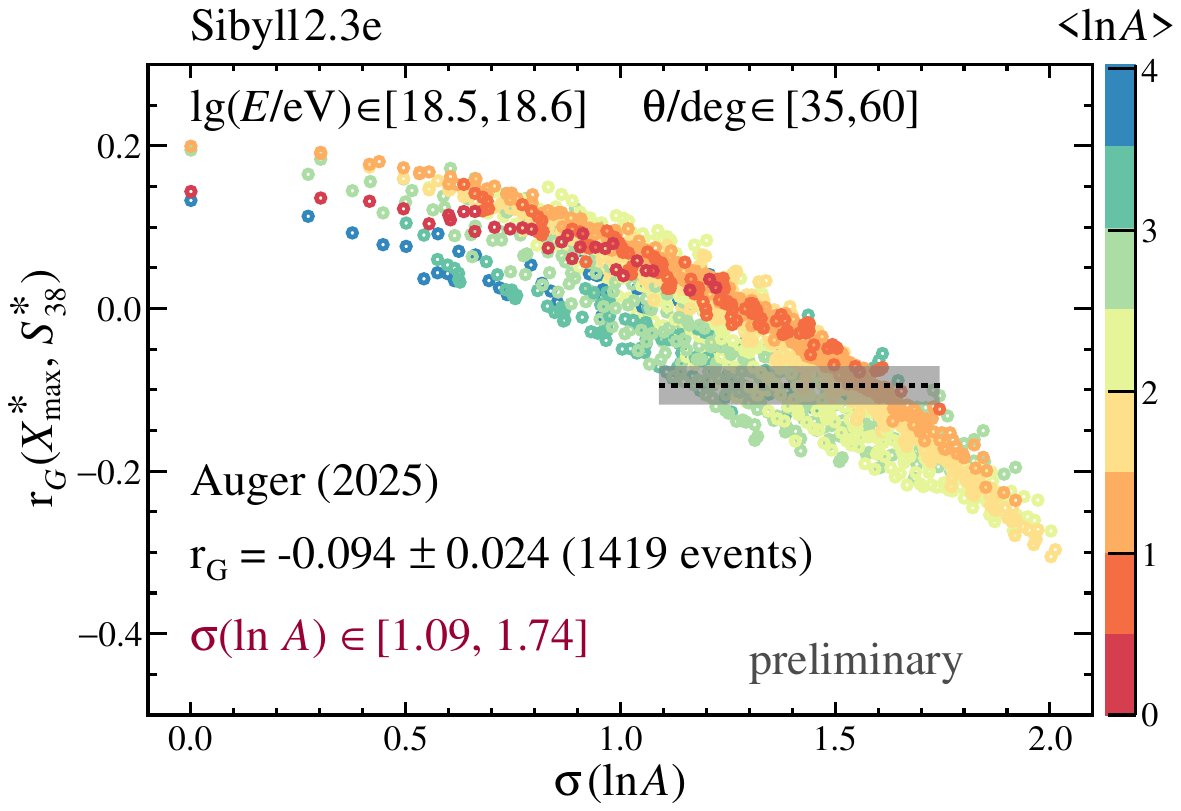}
  \includegraphics[width=0.49\textwidth]{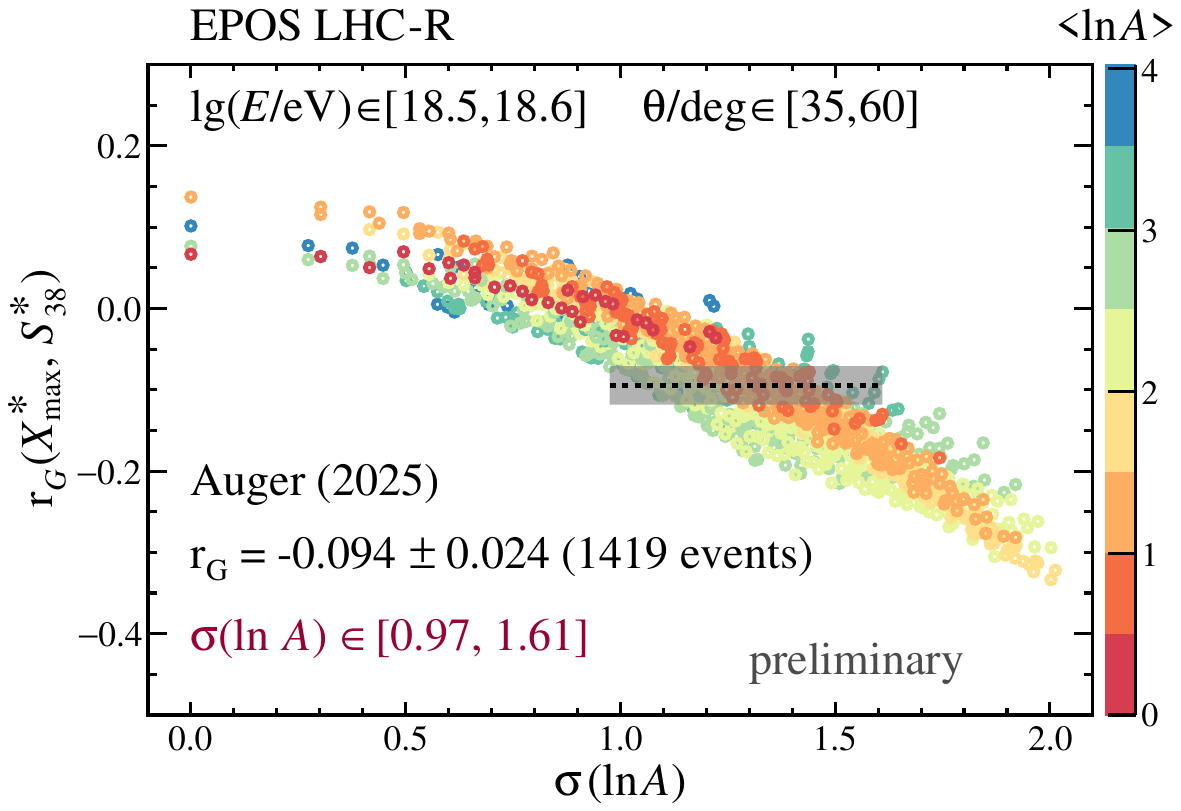}
  \caption{Conversion from \rG to \sigmalnA using simulations with
    \sibylle (left) and \eposlhcr (right). The shaded area represents
    the observed value with its statistical errors. The \sigmalnA ranges
    of the simulated mixes compatible with the correlation in the data
    are shown in each panel. See text for more details.}
  \label{fig:rg_to_sigmalna}
\end{figure}

\begin{figure}
  \centering
  \includegraphics[width=0.49\textwidth]{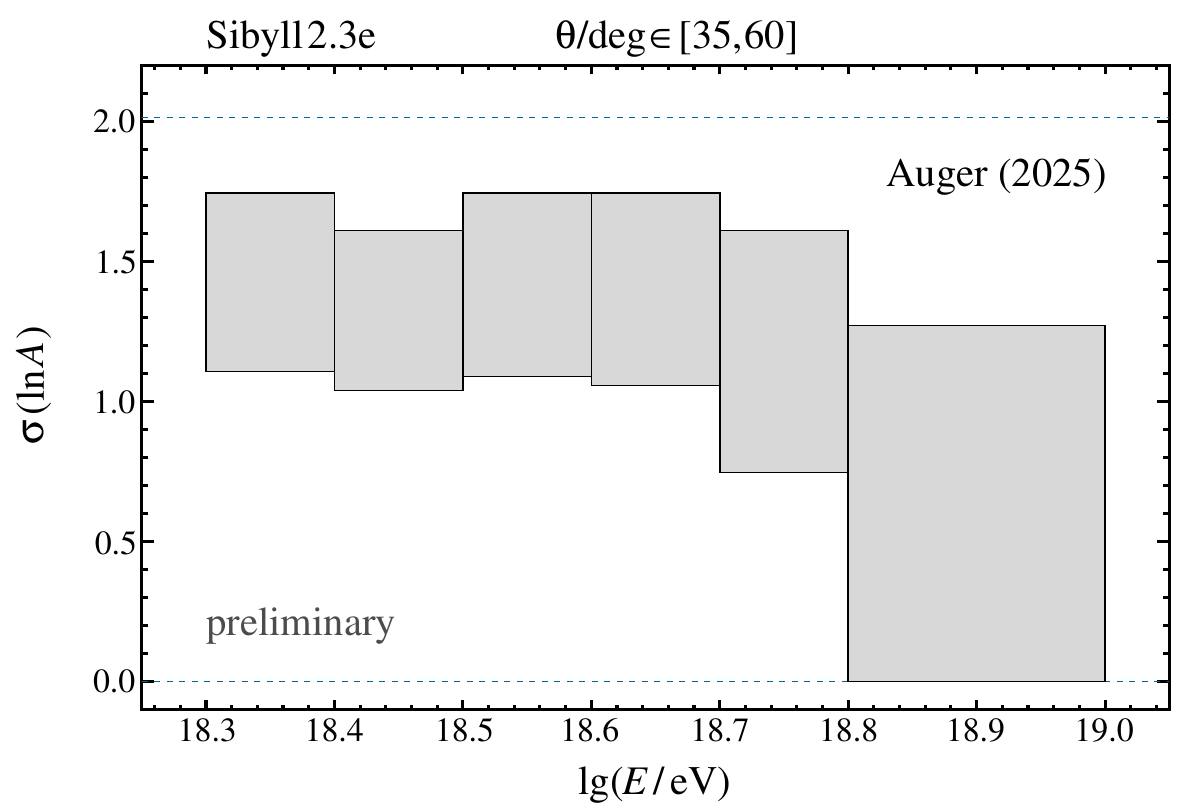}
    \includegraphics[width=0.49\textwidth]{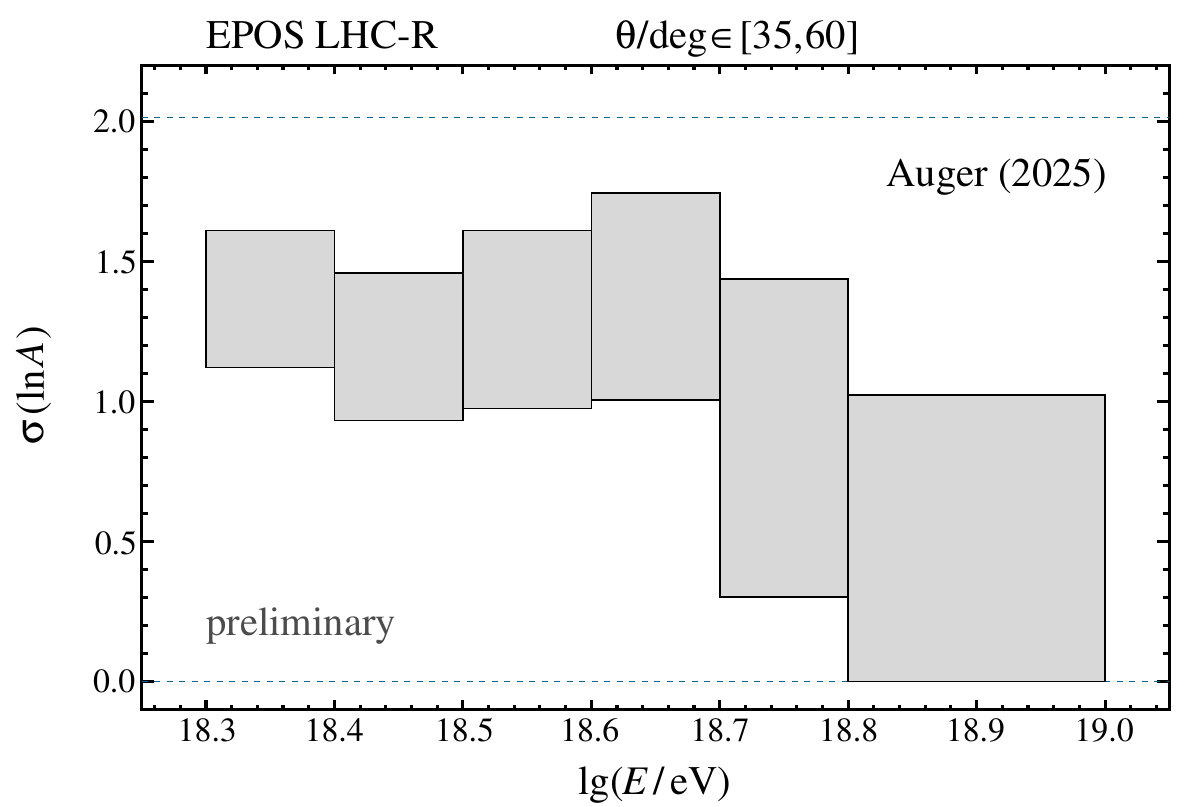}
  \caption{Constraints on \sigmalnA{} from the correlation analysis as
    a function of energy using \sibylle and \eposlhcr. Dashed
    lines indicate the values for pure compositions and the
    proton-iron equal mix (maximum mixing degree).}
    \label{fig:sigmalna_energy}
\end{figure}

To interpret these findings one needs to convert \rG to the spread of
the masses, \sigmalnA, using~\cref{fig:rg_to_sigmalna}. Each simulated
point in this plot corresponds to a mixture with different fractions
of (p, He, O, Fe) nuclei, the relative fractions change in 0.05 steps
(4 points for pure compositions are grouped at $\sigmalnA=0$). Colors
of the points indicate \meanlnA of each mix. In the energy bin
\lgerange{18.5}{18.6}, shown in this plot, the \rG value in the data
is compatible with $\sigmalnA\in[1.0,1.7]$. The evolution of the
spread of the masses for the energy range \lgerange{18.3}{19.0},
obtained using this approach, is presented
in~\cref{fig:sigmalna_energy}. Independently of the hadronic interaction model
used, in the energy range \lgerange{18.3}{18.7}, the spread of the
masses remains within $\sigmalnA\in[1.0,1.7]$ interval. At higher
energies, the limits on the mixing become $\sigmalnA\in[0.0,1.3]$
(also for \lgerange{19.0}{19.5}, not shown here). These conclusions
are valid for previous and even pre-LHC versions of the considered
hadronic interaction models~\cite{auger_mixed_plb2016}.

In~\cref{fig:rg_energy}, the discrepancy with the observed
correlation below the ankle for \sibylle can be explained by the
proton-helium dominated FD \Xmax mix with typically $<15\%$
contribution of nitrogen and $\sigmalnA\in[0.8,1.0]$. For \eposlhcr,
instead, the FD \Xmax mix is dominated by protons and nitrogen (helium
fraction is at maximum $\sim20\%$) resulting in a larger mixing degree
of $\sigmalnA\in[1.2,1.4]$ compatible with the constraints from the
\rG analysis. Let us note that, in the forthcoming publication, the
statistical and systematic uncertainties associated with the FD \Xmax
mix fractions will be propagated to the \rG values to properly
quantify these observations.

\begin{figure}
  \centering
  \includegraphics[width=0.49\textwidth]{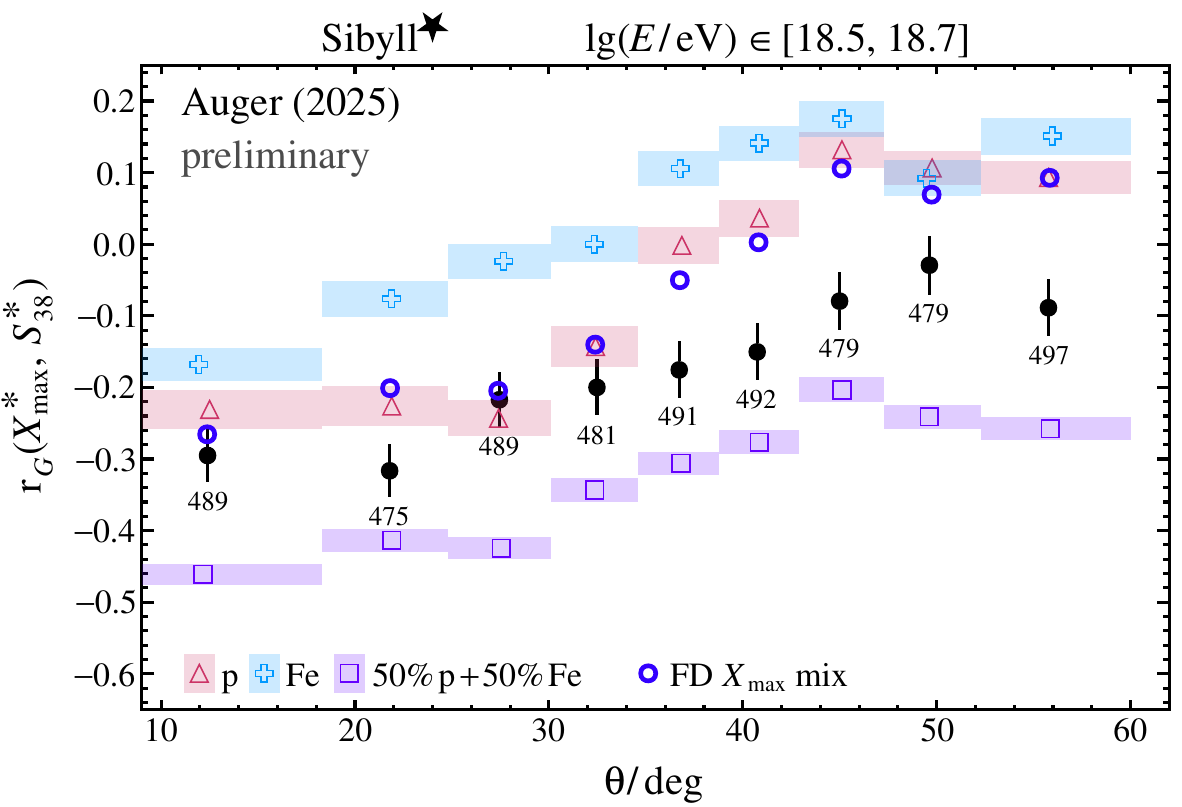}
    \includegraphics[width=0.49\textwidth]{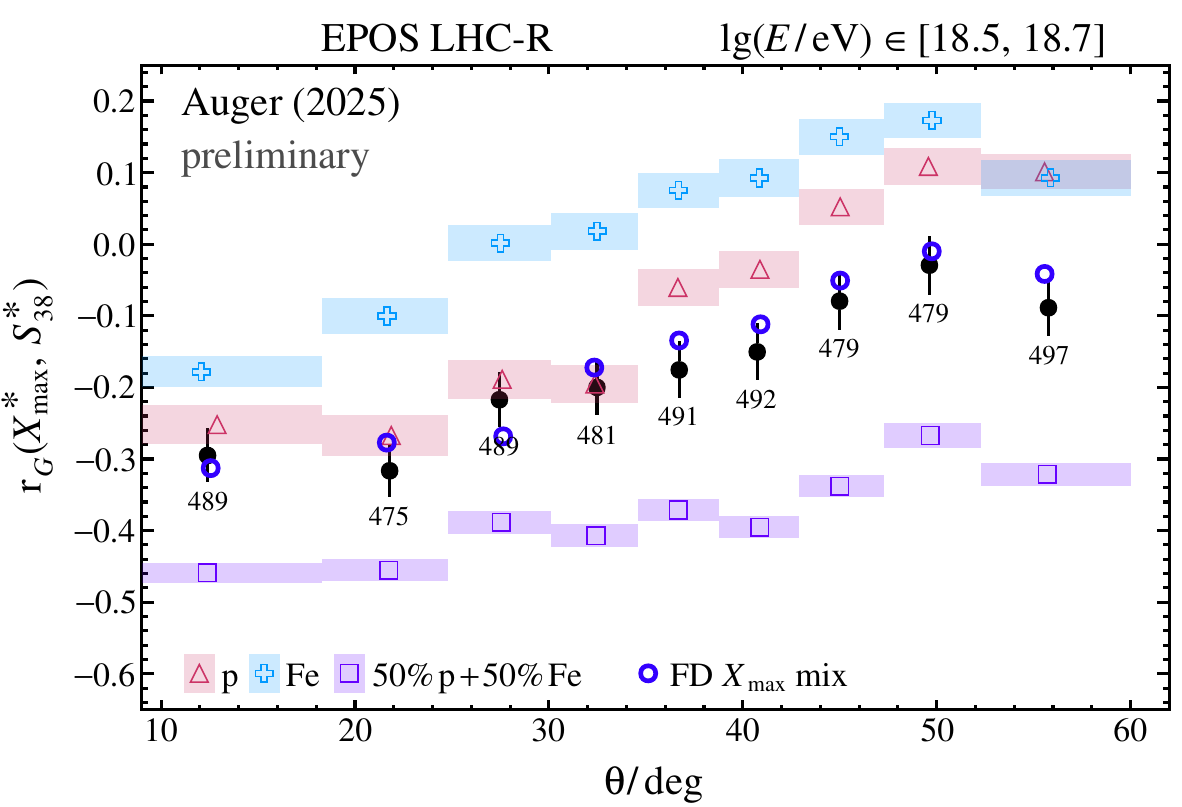}
  \caption{Zenith-angle dependence of \rG. The \rG values in the data
    are compared with \sibyllstar (left) and \eposlhcr (right)
    simulations for protons, iron nuclei, proton-iron equal mix, and
    the FD \Xmax mix. Energy range: \lgerange{18.5}{18.7}.}
  \label{fig:rg_zenith}
\end{figure}

\section{Zenith-angle dependence of the correlation}
\label{sec:zenith}

In \cref{fig:rg_zenith}, the zenith-angle dependence of the observed
correlation for the energy range \lgerange{18.5}{18.7} is compared with
the hadronic interaction model predictions for (proton, iron, proton-iron equal
mix) compositions and FD \Xmax mixes for \sibyllstar and
\eposlhcr. These hadronic interaction models represent the extreme (currently
existing) cases with the maximum muon content and the deepest \Xmax
scale respectively. However, the qualitative behavior of $\rG(\theta)$
is similar across all other models. Regarding simulations, the \rG
values for pure beams, particularly protons, remain significantly
negative up to a zenith angle of \ang{35}. This anticorrelation does
not become stronger even for relatively large degrees of mixing, as
evidenced by the similarity of the correlations for pure protons, the
FD \Xmax mix, and the data. As a consequence, the difference between
\rG for pure protons and the extreme mix is approximately two times
smaller than for zenith angles above \ang{35}. Exclusion of the events
below \ang{35} also improves the \sigmalnA constraints by reducing the
width of the simulated distributions shown
in~\cref{fig:rg_to_sigmalna}, as it makes the \rG values for pure
beams more similar to one another.

For zenith angles above \ang{35}, the discrepancies between the
correlations in the data and those for the FD \Xmax mix are evident in
the case of \sibyllstar. This demonstrates that the $\sim30\%$ ad-hoc
increase in the muon shower content in this model had no substantial
impact on the simulated correlation values, and \rG for the FD \Xmax
mix remained incompatible with the observed correlation, as was the
case for \sibylle (c.f. \cref{fig:rg_energy}).

In contrast, for \eposlhcr, its deeper \Xmax scale results in a larger
spread of masses in the FD \Xmax mix, leading to a correlation similar
to that observed in the data. Similar conclusions were obtained
in~\cite{auger_xmaxsdsignal_scale_prd2024}, where a good description
of the observed \SdSizeKm and \Xmax distributions was achieved by
increasing the hadronic signal and shifting \Xmax to deeper values in
earlier versions of the hadronic interaction models used here. The
shift of the \Xmax scale in~\cite{auger_xmaxsdsignal_scale_prd2024}
led to inferences with a larger spread of masses and an improved
description of the zenith-angle dependence of the correlation.

\section{Summary}

In this work, we have established constraints nearly independent of
hadronic interaction models on the spread of the UHECR masses
using the correlation between \Xmax and \SdSizeKm for events with
energies \lgerange{18.3}{19.5}. For energies below the ankle in the
UHECR spectrum, where we find $\sigmalnA\in[1.0,1.7]$, the mixes
consisting only of the neighboring mass groups (p-He, He-CNO, CNO-Fe)
are excluded. For higher energies, the mixing degree decreases to
$\sigmalnA\in[0.0,1.3]$. These conclusions hold for any hadronic
interaction model, including their pre-LHC versions, despite the fact
that differences between models in muon shower content exceed the
differences between proton and iron, and the differences in \Xmax
scale amount to about half the proton-iron distance.

By analyzing the energy and zenith-angle dependences of the observed
correlation for energies below the ankle, we established that the mass
compositions obtained from the fits of the FD \Xmax distributions are
not compatible with the correlation in the data for all interaction
models, except for \eposlhcr. The deeper \Xmax scale in \eposlhcr,
compared with other hadronic interaction models, results in FD \Xmax mixes with the
spread of the masses $\sigmalnA\in[1.2,1.4]$, with the energy and
zenith-angle dependencies of the correlation agreeing with the
observed ones. Similar results were obtained in our
analysis~\cite{auger_xmaxsdsignal_scale_prd2024}, which employed
previous versions of the hadronic interaction models with their \Xmax
scales shifted to deeper values, resembling the \Xmax scale of
\eposlhcr. Therefore, to match the observed correlation, adjusting the
\Xmax scale must result in a stronger mass mixing in the compositions
inferred from the \Xmax data. In addition, to achieve the same
outcome, modifications to the \Xmax fluctuations or the proton-iron
difference in \meanXmax{} could be considered.

A more detailed examination of the implications of the correlation
analysis for the UHECR mass composition and the validation of the
hadronic interaction model predictions will be reported in our forthcoming
publication.

\input{Yushkov_MASS2_CorrelationXmaxS38_ICRC2025.bbl}

\newpage

\section*{The Pierre Auger Collaboration}

{\footnotesize\setlength{\baselineskip}{11pt}
\noindent
\begin{wrapfigure}[11]{l}{0.12\linewidth}
\vspace{-4pt}
\includegraphics[width=0.98\linewidth]{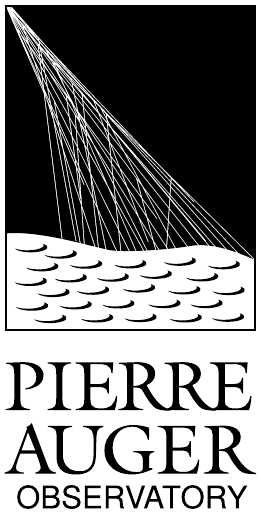}
\end{wrapfigure}
\begin{sloppypar}\noindent
\input{latex_authorlist_authors}
\end{sloppypar}
\begin{center}
\end{center}

\vspace{1ex}
\input{latex_authorlist_institutions}

\input{acknowledgments}
}

\end{document}

%% file: Yushkov_MASS2_CorrelationXmaxS38_ICRC2025.bbl
\providecommand{\href}[2]{#2}\begingroup\raggedright\endgroup

%% file: latex_authorlist_authors.tex
A.~Abdul Halim$^{13}$,
P.~Abreu$^{70}$,
M.~Aglietta$^{53,51}$,
I.~Allekotte$^{1}$,
K.~Almeida Cheminant$^{78,77}$,
A.~Almela$^{7,12}$,
R.~Aloisio$^{44,45}$,
J.~Alvarez-Mu\~niz$^{76}$,
A.~Ambrosone$^{44}$,
J.~Ammerman Yebra$^{76}$
% ...

%% file: latex_authorlist_institutions.tex
\begin{description}[labelsep=0.2em,align=right,labelwidth=0.7em,labelindent=0em,leftmargin=2em,noitemsep,before={\renewcommand\makelabel[1]{##1 }}]
\item[$^{1}$] Centro At\'omico Bariloche and Instituto Balseiro (CNEA-UNCuyo-CONICET), San Carlos de Bariloche, Argentina
\item[$^{2}$] Departamento de F\'\i{}sica and Departamento de Ciencias de la Atm\'osfera y los Oc\'eanos, FCEyN, Universidad de Buenos Aires and CONICET, Buenos Aires, Argentina
\item[$^{3}$] IFLP, Universidad Nacional de La Plata and CONICET, La Plata, Argentina
\item[$^{4}$] Instituto de Astronom\'\i{}a y F\'\i{}sica del Espacio (IAFE, CONICET-UBA), Buenos Aires, Argentina
\item[$^{5}$] Instituto de F\'\i{}sica de Rosario (IFIR) -- CONICET/U.N.R.\ and Facultad de Ciencias Bioqu\'\i{}micas y Farmac\'euticas U.N.R., Rosario, Argentina
\item[$^{6}$] Instituto de Tecnolog\'\i{}as en Detecci\'on y Astropart\'\i{}culas (CNEA, CONICET, UNSAM), and Universidad Tecnol\'ogica Nacional -- Facultad Regional Mendoza (CONICET/CNEA), Mendoza, Argentina
\item[$^{7}$] Instituto de Tecnolog\'\i{}as en Detecci\'on y Astropart\'\i{}culas (CNEA, CONICET, UNSAM), Buenos Aires, Argentina
\item[$^{8}$] International Center of Advanced Studies and Instituto de Ciencias F\'\i{}sicas, ECyT-UNSAM and CONICET, Campus Miguelete -- San Mart\'\i{}n, Buenos Aires, Argentina
\item[$^{9}$] Laboratorio Atm\'osfera -- Departamento de Investigaciones en L\'aseres y sus Aplicaciones -- UNIDEF (CITEDEF-CONICET), Argentina
\item[$^{10}$] Observatorio Pierre Auger, Malarg\"ue, Argentina
\item[$^{11}$] Observatorio Pierre Auger and Comisi\'on Nacional de Energ\'\i{}a At\'omica, Malarg\"ue, Argentina
\item[$^{12}$] Universidad Tecnol\'ogica Nacional -- Facultad Regional Buenos Aires, Buenos Aires, Argentina
\item[$^{13}$] University of Adelaide, Adelaide, S.A., Australia
\item[$^{14}$] Universit\'e Libre de Bruxelles (ULB), Brussels, Belgium
\item[$^{15}$] Vrije Universiteit Brussels, Brussels, Belgium
\item[$^{16}$] Centro Brasileiro de Pesquisas Fisicas, Rio de Janeiro, RJ, Brazil
\item[$^{17}$] Centro Federal de Educa\c{c}\~ao Tecnol\'ogica Celso Suckow da Fonseca, Petropolis, Brazil
\item[$^{18}$] Instituto Federal de Educa\c{c}\~ao, Ci\^encia e Tecnologia do Rio de Janeiro (IFRJ), Brazil
\item[$^{19}$] Universidade de S\~ao Paulo, Escola de Engenharia de Lorena, Lorena, SP, Brazil
\item[$^{20}$] Universidade de S\~ao Paulo, Instituto de F\'\i{}sica de S\~ao Carlos, S\~ao Carlos, SP, Brazil
\item[$^{21}$] Universidade de S\~ao Paulo, Instituto de F\'\i{}sica, S\~ao Paulo, SP, Brazil
\item[$^{22}$] Universidade Estadual de Campinas (UNICAMP), IFGW, Campinas, SP, Brazil
\item[$^{23}$] Universidade Estadual de Feira de Santana, Feira de Santana, Brazil
\item[$^{24}$] Universidade Federal de Campina Grande, Centro de Ciencias e Tecnologia, Campina Grande, Brazil
\item[$^{25}$] Universidade Federal do ABC, Santo Andr\'e, SP, Brazil
\item[$^{26}$] Universidade Federal do Paran\'a, Setor Palotina, Palotina, Brazil
\item[$^{27}$] Universidade Federal do Rio de Janeiro, Instituto de F\'\i{}sica, Rio de Janeiro, RJ, Brazil
\item[$^{28}$] Universidad de Medell\'\i{}n, Medell\'\i{}n, Colombia
\item[$^{29}$] Universidad Industrial de Santander, Bucaramanga, Colombia
\item[$^{30}$] Charles University, Faculty of Mathematics and Physics, Institute of Particle and Nuclear Physics, Prague, Czech Republic
\item[$^{31}$] Institute of Physics of the Czech Academy of Sciences, Prague, Czech Republic
\item[$^{32}$] Palacky University, Olomouc, Czech Republic
\item[$^{33}$] CNRS/IN2P3, IJCLab, Universit\'e Paris-Saclay, Orsay, France
\item[$^{34}$] Laboratoire de Physique Nucl\'eaire et de Hautes Energies (LPNHE), Sorbonne Universit\'e, Universit\'e de Paris, CNRS-IN2P3, Paris, France
\item[$^{35}$] Univ.\ Grenoble Alpes, CNRS, Grenoble Institute of Engineering Univ.\ Grenoble Alpes, LPSC-IN2P3, 38000 Grenoble, France
\item[$^{36}$] Universit\'e Paris-Saclay, CNRS/IN2P3, IJCLab, Orsay, France
\item[$^{37}$] Bergische Universit\"at Wuppertal, Department of Physics, Wuppertal, Germany
\item[$^{38}$] Karlsruhe Institute of Technology (KIT), Institute for Experimental Particle Physics, Karlsruhe, Germany
\item[$^{39}$] Karlsruhe Institute of Technology (KIT), Institut f\"ur Prozessdatenverarbeitung und Elektronik, Karlsruhe, Germany
\item[$^{40}$] Karlsruhe Institute of Technology (KIT), Institute for Astroparticle Physics, Karlsruhe, Germany
\item[$^{41}$] RWTH Aachen University, III.\ Physikalisches Institut A, Aachen, Germany
\item[$^{42}$] Universit\"at Hamburg, II.\ Institut f\"ur Theoretische Physik, Hamburg, Germany
\item[$^{43}$] Universit\"at Siegen, Department Physik -- Experimentelle Teilchenphysik, Siegen, Germany
\item[$^{44}$] Gran Sasso Science Institute, L'Aquila, Italy
\item[$^{45}$] INFN Laboratori Nazionali del Gran Sasso, Assergi (L'Aquila), Italy
\item[$^{46}$] INFN, Sezione di Catania, Catania, Italy
\item[$^{47}$] INFN, Sezione di Lecce, Lecce, Italy
\item[$^{48}$] INFN, Sezione di Milano, Milano, Italy
\item[$^{49}$] INFN, Sezione di Napoli, Napoli, Italy
\item[$^{50}$] INFN, Sezione di Roma ``Tor Vergata'', Roma, Italy
\item[$^{51}$] INFN, Sezione di Torino, Torino, Italy
\item[$^{52}$] Istituto di Astrofisica Spaziale e Fisica Cosmica di Palermo (INAF), Palermo, Italy
\item[$^{53}$] Osservatorio Astrofisico di Torino (INAF), Torino, Italy
\item[$^{54}$] Politecnico di Milano, Dipartimento di Scienze e Tecnologie Aerospaziali , Milano, Italy
\item[$^{55}$] Universit\`a del Salento, Dipartimento di Matematica e Fisica ``E.\ De Giorgi'', Lecce, Italy
\item[$^{56}$] Universit\`a dell'Aquila, Dipartimento di Scienze Fisiche e Chimiche, L'Aquila, Italy
\item[$^{57}$] Universit\`a di Catania, Dipartimento di Fisica e Astronomia ``Ettore Majorana``, Catania, Italy
\item[$^{58}$] Universit\`a di Milano, Dipartimento di Fisica, Milano, Italy
\item[$^{59}$] Universit\`a di Napoli ``Federico II'', Dipartimento di Fisica ``Ettore Pancini'', Napoli, Italy
\item[$^{60}$] Universit\`a di Palermo, Dipartimento di Fisica e Chimica ''E.\ Segr\`e'', Palermo, Italy
\item[$^{61}$] Universit\`a di Roma ``Tor Vergata'', Dipartimento di Fisica, Roma, Italy
\item[$^{62}$] Universit\`a Torino, Dipartimento di Fisica, Torino, Italy
\item[$^{63}$] Benem\'erita Universidad Aut\'onoma de Puebla, Puebla, M\'exico
\item[$^{64}$] Unidad Profesional Interdisciplinaria en Ingenier\'\i{}a y Tecnolog\'\i{}as Avanzadas del Instituto Polit\'ecnico Nacional (UPIITA-IPN), M\'exico, D.F., M\'exico
\item[$^{65}$] Universidad Aut\'onoma de Chiapas, Tuxtla Guti\'errez, Chiapas, M\'exico
\item[$^{66}$] Universidad Michoacana de San Nicol\'as de Hidalgo, Morelia, Michoac\'an, M\'exico
\item[$^{67}$] Universidad Nacional Aut\'onoma de M\'exico, M\'exico, D.F., M\'exico
\item[$^{68}$] Institute of Nuclear Physics PAN, Krakow, Poland
\item[$^{69}$] University of \L{}\'od\'z, Faculty of High-Energy Astrophysics,\L{}\'od\'z, Poland
\item[$^{70}$] Laborat\'orio de Instrumenta\c{c}\~ao e F\'\i{}sica Experimental de Part\'\i{}culas -- LIP and Instituto Superior T\'ecnico -- IST, Universidade de Lisboa -- UL, Lisboa, Portugal
\item[$^{71}$] ``Horia Hulubei'' National Institute for Physics and Nuclear Engineering, Bucharest-Magurele, Romania
\item[$^{72}$] Institute of Space Science, Bucharest-Magurele, Romania
\item[$^{73}$] Center for Astrophysics and Cosmology (CAC), University of Nova Gorica, Nova Gorica, Slovenia
\item[$^{74}$] Experimental Particle Physics Department, J.\ Stefan Institute, Ljubljana, Slovenia
\item[$^{75}$] Universidad de Granada and C.A.F.P.E., Granada, Spain
\item[$^{76}$] Instituto Galego de F\'\i{}sica de Altas Enerx\'\i{}as (IGFAE), Universidade de Santiago de Compostela, Santiago de Compostela, Spain
\item[$^{77}$] IMAPP, Radboud University Nijmegen, Nijmegen, The Netherlands
\item[$^{78}$] Nationaal Instituut voor Kernfysica en Hoge Energie Fysica (NIKHEF), Science Park, Amsterdam, The Netherlands
\item[$^{79}$] Stichting Astronomisch Onderzoek in Nederland (ASTRON), Dwingeloo, The Netherlands
\item[$^{80}$] Universiteit van Amsterdam, Faculty of Science, Amsterdam, The Netherlands
\item[$^{81}$] Case Western Reserve University, Cleveland, OH, USA
\item[$^{82}$] Colorado School of Mines, Golden, CO, USA
\item[$^{83}$] Department of Physics and Astronomy, Lehman College, City University of New York, Bronx, NY, USA
\item[$^{84}$] Michigan Technological University, Houghton, MI, USA
\item[$^{85}$] New York University, New York, NY, USA
\item[$^{86}$] University of Chicago, Enrico Fermi Institute, Chicago, IL, USA
\item[$^{87}$] University of Delaware, Department of Physics and Astronomy, Bartol Research Institute, Newark, DE, USA
\item[] -----
\item[$^{a}$] Max-Planck-Institut f\"ur Radioastronomie, Bonn, Germany
\item[$^{b}$] also at Kapteyn Institute, University of Groningen, Groningen, The Netherlands
\item[$^{c}$] School of Physics and Astronomy, University of Leeds, Leeds, United Kingdom
\item[$^{d}$] Fermi National Accelerator Laboratory, Fermilab, Batavia, IL, USA
\item[$^{e}$] Pennsylvania State University, University Park, PA, USA
\item[$^{f}$] Colorado State University, Fort Collins, CO, USA
\item[$^{g}$] Louisiana State University, Baton Rouge, LA, USA
\item[$^{h}$] now at Graduate School of Science, Osaka Metropolitan University, Osaka, Japan
\item[$^{i}$] Institut universitaire de France (IUF), France
\item[$^{j}$] now at Technische Universit\"at Dortmund and Ruhr-Universit\"at Bochum, Dortmund and Bochum, Germany
\end{description}

%% file: acknowledgments.tex
\section*{Acknowledgments}

\begin{sloppypar}
The successful installation, commissioning, and operation of the Pierre
Auger Observatory would not have been possible without the strong
commitment and effort from the technical and administrative staff in
Malarg\"ue. We are very grateful to the following agencies and
organizations for financial support:
\end{sloppypar}

\begin{sloppypar}
Argentina -- Comisi\'on Nacional de Energ\'\i{}a At\'omica; Agencia Nacional de
Promoci\'on Cient\'\i{}fica y Tecnol\'ogica (ANPCyT); Consejo Nacional de
Investigaciones Cient\'\i{}ficas y T\'ecnicas (CONICET); Gobierno de la
Provincia de Mendoza; Municipalidad de Malarg\"ue; NDM Holdings and Valle
Las Le\~nas; in gratitude for their continuing cooperation over land
access; Australia -- the Australian Research Council; Belgium -- Fonds
de la Recherche Scientifique (FNRS); Research Foundation Flanders (FWO),
Marie Curie Action of the European Union Grant No.~101107047; Brazil --
Conselho Nacional de Desenvolvimento Cient\'\i{}fico e Tecnol\'ogico (CNPq);
Financiadora de Estudos e Projetos (FINEP); Funda\c{c}\~ao de Amparo \`a
Pesquisa do Estado de Rio de Janeiro (FAPERJ); S\~ao Paulo Research
Foundation (FAPESP) Grants No.~2019/10151-2, No.~2010/07359-6 and
No.~1999/05404-3; Minist\'erio da Ci\^encia, Tecnologia, Inova\c{c}\~oes e
Comunica\c{c}\~oes (MCTIC); Czech Republic -- GACR 24-13049S, CAS LQ100102401,
MEYS LM2023032, CZ.02.1.01/0.0/0.0/16{\textunderscore}013/0001402,
CZ.02.1.01/0.0/0.0/18{\textunderscore}046/0016010 and
CZ.02.1.01/0.0/0.0/17{\textunderscore}049/0008422 and CZ.02.01.01/00/22{\textunderscore}008/0004632;
France -- Centre de Calcul IN2P3/CNRS; Centre National de la Recherche
Scientifique (CNRS); Conseil R\'egional Ile-de-France; D\'epartement
Physique Nucl\'eaire et Corpusculaire (PNC-IN2P3/CNRS); D\'epartement
Sciences de l'Univers (SDU-INSU/CNRS); Institut Lagrange de Paris (ILP)
Grant No.~LABEX ANR-10-LABX-63 within the Investissements d'Avenir
Programme Grant No.~ANR-11-IDEX-0004-02; Germany -- Bundesministerium
f\"ur Bildung und Forschung (BMBF); Deutsche Forschungsgemeinschaft (DFG);
Finanzministerium Baden-W\"urttemberg; Helmholtz Alliance for
Astroparticle Physics (HAP); Helmholtz-Gemeinschaft Deutscher
Forschungszentren (HGF); Ministerium f\"ur Kultur und Wissenschaft des
Landes Nordrhein-Westfalen; Ministerium f\"ur Wissenschaft, Forschung und
Kunst des Landes Baden-W\"urttemberg; Italy -- Istituto Nazionale di
Fisica Nucleare (INFN); Istituto Nazionale di Astrofisica (INAF);
Ministero dell'Universit\`a e della Ricerca (MUR); CETEMPS Center of
Excellence; Ministero degli Affari Esteri (MAE), ICSC Centro Nazionale
di Ricerca in High Performance Computing, Big Data and Quantum
Computing, funded by European Union NextGenerationEU, reference code
CN{\textunderscore}00000013; M\'exico -- Consejo Nacional de Ciencia y Tecnolog\'\i{}a
(CONACYT) No.~167733; Universidad Nacional Aut\'onoma de M\'exico (UNAM);
PAPIIT DGAPA-UNAM; The Netherlands -- Ministry of Education, Culture and
Science; Netherlands Organisation for Scientific Research (NWO); Dutch
national e-infrastructure with the support of SURF Cooperative; Poland
-- Ministry of Education and Science, grants No.~DIR/WK/2018/11 and
2022/WK/12; National Science Centre, grants No.~2016/22/M/ST9/00198,
2016/23/B/ST9/01635, 2020/39/B/ST9/01398, and 2022/45/B/ST9/02163;
Portugal -- Portuguese national funds and FEDER funds within Programa
Operacional Factores de Competitividade through Funda\c{c}\~ao para a Ci\^encia
e a Tecnologia (COMPETE); Romania -- Ministry of Research, Innovation
and Digitization, CNCS-UEFISCDI, contract no.~30N/2023 under Romanian
National Core Program LAPLAS VII, grant no.~PN 23 21 01 02 and project
number PN-III-P1-1.1-TE-2021-0924/TE57/2022, within PNCDI III; Slovenia
-- Slovenian Research Agency, grants P1-0031, P1-0385, I0-0033, N1-0111;
Spain -- Ministerio de Ciencia e Innovaci\'on/Agencia Estatal de
Investigaci\'on (PID2019-105544GB-I00, PID2022-140510NB-I00 and
RYC2019-027017-I), Xunta de Galicia (CIGUS Network of Research Centers,
Consolidaci\'on 2021 GRC GI-2033, ED431C-2021/22 and ED431F-2022/15),
Junta de Andaluc\'\i{}a (SOMM17/6104/UGR and P18-FR-4314), and the European
Union (Marie Sklodowska-Curie 101065027 and ERDF); USA -- Department of
Energy, Contracts No.~DE-AC02-07CH11359, No.~DE-FR02-04ER41300,
No.~DE-FG02-99ER41107 and No.~DE-SC0011689; National Science Foundation,
Grant No.~0450696, and NSF-2013199; The Grainger Foundation; Marie
Curie-IRSES/EPLANET; European Particle Physics Latin American Network;
and UNESCO.
\end{sloppypar}

%% file: Yushkov_MASS2_CorrelationXmaxS38_ICRC2025_Id678_Submit_PoS442.bbl
\begin{thebibliography}{10}

\bibitem{younk_corr_app2012}
P.~Younk and M.~Risse, \emph{{Sensitivity of the correlation between the depth
  of shower maximum and the muon shower size to the cosmic ray composition}},
  \href{https://doi.org/10.1016/j.astropartphys.2012.03.001}{\emph{Astropart.Phys.}
  {\bfseries 35} (2012) 807} [\href{https://arxiv.org/abs/1203.3732}{{\ttfamily
  1203.3732}}].

\bibitem{auger_nima2015}
{\scshape Pierre Auger Collaboration}, \emph{{The Pierre Auger Cosmic Ray
  Observatory}}, \href{https://doi.org/10.1016/j.nima.2015.06.058}{\emph{Nucl.
  Instrum. Meth.} {\bfseries A798} (2015) 172}
  [\href{https://arxiv.org/abs/1502.01323}{{\ttfamily 1502.01323}}].

\bibitem{auger_mixed_plb2016}
{\scshape Pierre Auger Collaboration}, \emph{{Evidence for a mixed mass
  composition at the `ankle' in the cosmic-ray spectrum}},
  \href{https://doi.org/10.1016/j.physletb.2016.09.039}{\emph{Phys. Lett.}
  {\bfseries B762} (2016) 288}
  [\href{https://arxiv.org/abs/1609.08567}{{\ttfamily 1609.08567}}].

\bibitem{fitoussi_pos2023}
T.~Fitoussi, \emph{{Depth of maximum of air-shower profiles above $10^{17.8}$
  eV measured with the fluorescence detector of the Pierre Auger Observatory
  and mass-composition implications}},
  \href{https://doi.org/10.22323/1.444.0249}{\emph{PoS} {\bfseries ICRC2023}
  (2023) 319}.

\bibitem{sdreco_jinst2021}
{\scshape Pierre Auger Collaboration}, \emph{{Reconstruction of events recorded
  with the surface detector of the Pierre Auger Observatory}},
  \href{https://doi.org/10.1088/1748-0221/15/10/P10021}{\emph{JINST} {\bfseries
  15} (2020) P10021} [\href{https://arxiv.org/abs/2007.09035}{{\ttfamily
  2007.09035}}].

\bibitem{santos_augerlib_icrc2023}
E. Santos, \emph{{Update on the Offline Framework
  for AugerPrime and production of reference simulation libraries using the VO
  Auger grid resources}}, \href{https://doi.org/10.22323/1.444.0248}{\emph{PoS}
  {\bfseries ICRC2023} (2023) 248}.

\bibitem{corsika}
D.~Heck et~al., \emph{{CORSIKA: A Monte Carlo code to simulate extensive air
  showers}}, Report FZKA-6019 (1998).

\bibitem{offline}
{\scshape Pierre Auger Collaboration}, \emph{{The Offline Software Framework of
  the Pierre Auger Observatory}},
  \href{https://doi.org/10.1016/j.nima.2007.07.010}{\emph{Nucl. Instrum.
  Methods Phys. Res.} {\bfseries 580} (2007) 1485}.

\bibitem{eposlhcr_pos2023}
T.~Pierog and K.~Werner, \emph{{EPOS LHC-R : up-to-date hadronic model for EAS
  simulations}}, \href{https://doi.org/10.22323/1.444.0230}{\emph{PoS}
  {\bfseries ICRC2023} (2023) 230}.

\bibitem{qgsjet3_i_prd2024}
S.~Ostapchenko, \emph{{QGSJET-III model of high energy hadronic interactions:
  The formalism}},
  \href{https://doi.org/10.1103/PhysRevD.109.034002}{\emph{Phys. Rev. D}
  {\bfseries 109} (2024) 034002}.

\bibitem{qgsjet3_ii_prd2024}
S.~Ostapchenko, \emph{{QGSJET-III model of high energy hadronic interactions.
  II. Particle production and extensive air shower characteristics}},
  \href{https://doi.org/10.1103/PhysRevD.109.094019}{\emph{Phys. Rev. D}
  {\bfseries 109} (2024) 094019}.

\bibitem{sibyllstar_icrc23}
F.~Riehn, R.~Engel and A.~Fedynitch, \emph{{Sibyll\,$^\bigstar$: ad-hoc
  modifications for an improved description of muon data in extensive air
  showers}}, \href{https://doi.org/10.22323/1.444.0429}{\emph{PoS} {\bfseries
  ICRC2023} (2023) 429} [\href{https://arxiv.org/abs/2309.05390}{{\ttfamily
  2309.05390}}].

\bibitem{sibyll23d_prd2020}
F.~Riehn et~al., \emph{{Hadronic interaction model Sibyll~2.3d and extensive
  air showers}}, \href{https://doi.org/10.1103/PhysRevD.102.063002}{\emph{Phys.
  Rev. D} {\bfseries 102} (2020) 063002}
  [\href{https://arxiv.org/abs/1912.03300}{{\ttfamily 1912.03300}}].

\bibitem{pierog_fzu2025} T.~Pierog, \emph{{EPOS LHC-R: A Global
  Approach to Solve the Muon Puzzle}}, Presented at FZU -- Institute of
Physics of the Czech Academy of Sciences (2025),
\url{https://indico.fzu.cz/event/303/}.

\bibitem{auger_sd1500_spectrum_prd2020}
{\scshape Pierre Auger Collaboration}, \emph{{Measurement of the cosmic-ray
  energy spectrum above $2.5{\times} 10^{18}$ eV using the Pierre Auger
  Observatory}}, \href{https://doi.org/10.1103/PhysRevD.102.062005}{\emph{Phys.
  Rev. D} {\bfseries 102} (2020) 062005}
  [\href{https://arxiv.org/abs/2008.06486}{{\ttfamily 2008.06486}}].

\bibitem{rg_gideon_jasa1987}
R.~Gideon and R.~Hollister, \emph{{A rank correlation coefficient resistant to
  outliers}},
  \href{https://doi.org/10.1080/01621459.1987.10478480}{\emph{Journal of the
  American Statistical Association} {\bfseries 82} (1987) 656}.

\bibitem{tkachenko_icrc2023}
O.~Tkachenko, \emph{{Studies of the mass composition
  of cosmic rays and proton-proton interaction cross-sections at ultra-high
  energies with the Pierre Auger Observatory}},
  \href{https://doi.org/10.22323/1.444.0438}{\emph{PoS} {\bfseries ICRC2023}
    (2023) 438}.

\bibitem{tkachenko_icrc2025} O.~Tkachenko, \emph{{Measurement of the
  Inelastic Proton-Proton Cross Section at $\sqrt{s}\geq 40$~TeV using
  the Hybrid Data of the Pierre Auger Observatory}},
  these proceedings.

\bibitem{auger_xmaxsdsignal_scale_prd2024}
{\scshape Pierre Auger Collaboration}, \emph{{Testing hadronic-model
  predictions of depth of maximum of air-shower profiles and ground-particle
  signals using hybrid data of the Pierre Auger Observatory}},
  \href{https://doi.org/10.1103/PhysRevD.109.102001}{\emph{Phys. Rev. D}
  {\bfseries 109} (2024) 102001}
  [\href{https://arxiv.org/abs/2401.10740}{{\ttfamily 2401.10740}}].

\end{thebibliography}
